\documentclass[11pt]{article}

\renewcommand{\baselinestretch}{1.2}
  \renewcommand{\arraystretch}{1.0}
\usepackage{cite}
\usepackage[square, comma, sort&compress, numbers]{natbib}
 \usepackage{amsmath}
\usepackage{amscd}
\usepackage{amssymb}
\usepackage{graphicx}
\usepackage{ifpdf}
\usepackage{fancyhdr}

\begin{document}

 \title{  A Note On Boneh-Gentry-Waters Broadcast \\ Encryption Scheme and Its Like}

  \author{Zhengjun Cao$^{1}$, \qquad Lihua Liu$^{2,*}$}
  \footnotetext{ $^1$Department of Mathematics, Shanghai University, Shanghai,
  China. \ \ $^2$Department of Mathematics, Shanghai Maritime University,   Shanghai,  China. \  $^*$\,\textsf{liulh@shmtu.edu.cn}}

\date{}
\maketitle

\begin{quotation} \textbf{Abstract}.
 Key establishment is any process whereby a shared secret key becomes available
to two or more parties, for subsequent cryptographic use such as symmetric-key encryption.
Though it is widely known that
the primitive of encryption is different from key establishment, we find some researchers have confused the two primitives. In this note, we shall clarify the fundamental difference between the two primitives, and
 point out that the Boneh-Gentry-Waters broadcast encryption scheme and its like
are key establishment schemes, not encryption schemes.

 \textbf{Keywords}. key establishment; encryption; broadcast encryption
 \end{quotation}

\section{Introduction}

Keeping information secret from all but those who are authorized
to see it, is a main objective of cryptography. The primitive of encryption can provide
the functionality. Over the centuries, a lot of  mechanisms have been created to
deal with the information security issue.

In 1976, Diffie and Hellman \cite{DH76} introduced the concept of public-key cryptography and also provided a new method for key establishment.
But they had not put forth any public-key encryption
scheme at the time. In 1978, Rivest, Shamir, and Adleman \cite{RSA78} discovered the first
practical public-key encryption and signature scheme which is based on the hard mathematical problem of factorization. To date, the computational performance of public-key encryption is inferior to that of symmetric-key encryption because of much larger working parameters needed.   So, public-key encryption schemes are generally used to establish a key for a symmetric-key system being used by communicating entities \cite{MOV96}.
That is to say, key establishment is really intertwined with encryption.
Then, what is the fundamental difference between key establishment and encryption?

We find all literatures have not specified the difference.
  We also find some researchers have confused key establishment and encryption. In this note, we shall clarify the difference between the two primitives, and
 point out that the Boneh-Gentry-Waters broadcast encryption scheme \cite{BGW05} and its like \cite{AFI06,DPP07,GW09,JK10,L13,L14,MM12}
are key establishment schemes, not encryption schemes.

\section{Key establishment}

\textbf{Definition 1} \cite{MOV96} \emph{Key establishment is any process whereby a shared secret key becomes available
to two or more parties, for subsequent cryptographic use. }

Key establishment can be broadly subdivided into key agreement and key transport. A \emph{key transport} protocol is a key establishment technique where
one party creates or otherwise obtains a secret value, and securely transfers it to the other(s). A \emph{key agreement} protocol is a key establishment technique in
which a shared secret is derived by two (or more) parties as a function of information contributed
by, or associated with, each of these, such that no party can predetermine
the resulting value. Key establishment protocols result in shared secrets which are typically used to
derive, \emph{session keys}. For example, the
 Diffie-Hellman key agreement scheme is such a protocol.
 \vspace*{2mm}

\begin{center} Table 1: Diffie-Hellman key agreement (basic version)\vspace*{2mm}

\begin{tabular}{|l|l|}
  \hline
    Setup&  A prime $p$ and generator $g$ of $Z_p^*$
 are selected and published. \\ \hline
 Protocol actions
 & (a) A picks a random $x, 1\leq x \leq p-2$, and sends $g^x \mod p$ to B.\\
& (b) B picks a random $y, 1\leq x \leq p-2$, and sends $g^y \mod p$ to A.\\
& (c) B computes the shared key as $K = (g^x)^y \mod p$.\\
& (d) A computes the shared key as $K = (g^y)^x \mod p$.\\
  \hline
  Result &  The shared secret $K$ is known to both parties A and B.\\ \hline
\end{tabular}

\end{center}
\vspace*{2mm}

 \section{Encryption}

\textbf{Definition 2} \cite{MOV96}  \emph{Let  $\mathcal{A}$ be a finite set called the alphabet of definition,
  $\mathcal{M}$ be a set called the message space, $\mathcal{C}$ be a set called the ciphertext space, $\mathcal{K}$ be a set called the key space.  Each element $e\in\mathcal{K}$ uniquely determines a bijection from $\mathcal{M}$ to $\mathcal{C}$, denoted by $E_e$. $E_e$ is called an encryption function or an encryption transformation. For each $d \in \mathcal{K}$, $D_d$ denotes a bijection from $\mathcal{C}$ to $\mathcal{M}$. $D_d$ is
called a decryption function or decryption transformation. An encryption scheme consists of a set $\{E_e : e \in K\}$  of encryption transformations and a corresponding set $\{D_d : d \in K\}$ of decryption transformations with the property
that for each $e \in \mathcal{K}$ there is a unique key $d \in \mathcal{K}$ such that $D_d(E_e(m))=m$ for all $m \in \mathcal{M}$. }

The above definition is somewhat tedious. We refer to  RSA system for a concrete example of encryption scheme, which is a well-known public-key encryption and signature scheme. Note that at the end of the scheme, both the sender and the intended receiver know the message. \vspace*{2mm}

 \begin{center}
Table 2:  RSA encryption\vspace*{2mm}

 \begin{tabular}{|l|l|}
   \hline
    Setup & Pick two distinct odd primes $p$ and $q$,  \\
  & compute $n=pq$, $\phi(n)=(p-1)(q-1)$.   \\
  & Pick $e\in \mathbb{Z}_n^*$, compute $d=e^{-1}\mod \phi(n)$  \\
  & Publish $n, e$ and keep $d, p, q$ in secret.
  \\ \hline
  Encrypting & For a message $m\in \mathbb{Z}_n$, compute $c=m^e\mod n.$ \\ \hline
   Decrypting & Use the secret key $d$ to recover $m=c^d \mod n.$ \\ \hline
   \end{tabular}
 \end{center}

  \section{A difference between key establishment and encryption}
  As we see, both Diffie-Hellman key agreement and RSA encryption can ensure the two participators to
  know a same thing, whether we call it a shared key or a message. Then, what are the  differences between key establishment and encryption? It is a pity that we find all literatures have not specified the differences.
  We think the fundamental difference between two primitives is that
\centerline{ \fbox{whether the resulting thing is pre-existing.}}
Concretely, in an encryption scheme, both two participators use the message as a whole. They do not use any components of the
    message for the related transformations. But in a key establishment scheme, at least one participator has to use some components of the shared key for related computations and the final composition. To illustrate this point, we refer to the following Table 3 for the difference between RSA encryption and Diffie-Hellman key agreement.

    \begin{center}
Table 3: The difference between  RSA encryption and Diffie-Hellman key agreement\vspace*{2mm}

 \begin{tabular}{|l|l|l|}\hline
& RSA encryption & Diffie-Hellman key agreement \\ \hline
Computation &     $c=m^e\mod n$, &  $g^x \mod p$,  $g^y \mod p$, \\
   & $m=c^d\mod n$.  & $K = (g^y)^x\mod p= (g^x)^y\mod p$. \\ \hline
   Result & Both the two parties know $m$.  &  Both the two parties know $K$. \\ \hline
   Characteristic &$m$ is pre-existing. & $K$ is not pre-existing.\\
   \hline
   \end{tabular}
 \end{center}

\section{The Boneh-Gentry-Waters ``broadcast encryption" scheme and its like are not true encryption schemes}

\subsection{The Boneh-Gentry-Waters ``broadcast encryption" scheme}
The primitive of broadcast encryption was formalized by Fiat and Naor \cite{FN93}, which requires that the broadcaster encrypts a message such that a particular
set of users can decrypt the message sent over a broadcast channel.
The Fiat-Naor broadcast encryption and the following works \cite{GSY99,GSW00,KRS99,S97,ST98} use a combinatorial
approach. This approach  has to right the  balance between the efficiency and the number of colluders that the system is resistant to.
Recently, Boneh et al \cite{BGW05} have constructed some ``broadcast encrypt" systems. In these systems, the public parameters must be updated to allow more users. But we find the Boneh-Gentry-Waters ``broadcast encryption" scheme and its like \cite{AFI06,DPP07,GW09,JK10,L13,L14,MM12} are not true encryption schemes. They are key establishment schemes. For convenience, we now relate the Boneh-Gentry-Waters scheme as follows.

\emph{Setup}($n$): Suppose there are $n$ users in the system. Let $\mathbb{G}$ be a bilinear group of prime order $p$. Pick a random generator
$g \in \mathbb{G}$ and  random  numbers $\alpha, \gamma \in \mathbb{Z}_p$. Compute $v=g^{\gamma}$ and $g_i = g^{(\alpha^i)}$ for $i =1, 2, \cdots, n, n+2, \cdots, 2n$. The public key is set as:
$$PK = (g, g_1, \cdots, g_n, g_{n+2}, \cdots, g_{2n}, v).$$
The private key for user $i\in\{1, \cdots, n\}$ is set as: $d_i = g_i^{\gamma}$.

\emph{Encrypt}($S, PK$): Let $S$ be the set of  the intending receivers. Pick a random $t\in \mathbb{Z}_p$ and set $K = e(g_{n+1}, g)^t$. The value $e(g_{n+1}, g)$ can
be computed as $e(g_n, g_1)$. Next, set
$$\mbox{Hdr}=\left(g^t, (v\cdot \prod_{j\in S}g_{n+1-j})^t\right)$$
and output the pair (Hdr, $K$).

\emph{Decrypt}($S, i, d_i, \mbox{Hdr}, PK$): Parse Hdr as $(C_0, C_1)$ and compute
$$ K = e(g_i, C_1)/e(d_i\cdot \prod_{\stackrel{j\in S}{j\neq i}}
g_{n+1-j+i}, C_0).$$

\subsection{Analysis} 

 It is easy to see that in the  Boneh-Gentry-Waters scheme the shared thing $K$ is not pre-existing. It depends on the choice of the encrypter. Moreover, the encrypter has to use its secret component $t$ for other computations. Of course, the scheme can be transformed into a regular encryption scheme. It only needs to
set $c=M\cdot e(g_{n+1}, g)^t$ for a given message $M\in \mathbb{G}$ in the Encryption phase. To recover
$M$, the  user with $d_i$ can compute
$$M=c\cdot e(d_i\cdot \prod_{\stackrel{j\in S}{j\neq i}}
g_{n+1-j+i}, C_0)/ e(g_i, C_1). $$
But we should remark that  pairings including Weil pairing and Tate pairing are derived from elliptic curves \cite{S86}. Both $K, M$ are in the extension field $\mathbb{F}_{q^{k}}$ where $\mathbb{F}_q$ is the base field of the elliptic curve defined over. $k$ is called the \emph{embedding degree} which is the smallest positive
integer such that $p$ divides $(q^k-1)$. That means the scheme has  to work in some extension
of the base field, even though the inputting parameters are defined over
the base field.  That is to say, the scheme has  to work in a running environment with parameters of 1024 bits, not 160 bits as supposed (someone has confused the \emph{inputting-parameter's size} with the \emph{working-parameter's size}),  so as to offer 80 bits security level.  The shortcoming makes the scheme lose its competitive advantages significantly.

\section{Conclusion}

We clarify the fundamental difference between  encryption and  key establishment. To the best of our knowledge, it is the first time to put forth such an explicit principle to discriminate the two primitives.  We also remark some schemes are not true encryption schemes, instead key establishment schemes.

\end{document}